\documentclass[12pt,preprint]{aastex}
\usepackage{natbib}
\usepackage{graphicx}
\usepackage{amssymb}
\usepackage{amsmath}
\shorttitle{Model-independent constraints on ultra-light dark matter}
\begin{document}
\title{Model-independent constraints on ultra-light dark matter from the SPARC data}
\author{Man Ho Chan, Chu Fai Yeung}
\affil{Department of Science and Environmental Studies, The Education University of Hong Kong, Hong Kong, China}
\email{chanmh@eduhk.hk}

\begin{abstract}
Ultra-light dark matter (ULDM) is currently one of the most popular classes of cosmological dark matter. The most important advantage is that ULDM with mass $m \sim 10^{-22}$ eV can account for the small-scale problems encountered in the standard cold dark matter (CDM) model like the core-cusp problem, missing satellite problem and the too-big-to-fail problem in galaxies. In this article, we formulate a new simple model-independent analysis using the SPARC data to constrain the range of ULDM mass. In particular, the most stringent constraint comes from the data of a galaxy ESO563-G021, which can conservatively exclude a ULDM mass range $m=(0.14-3.11)\times 10^{-22}$ eV. This model-independent excluded range is consistent with many bounds obtained by recent studies and it suggests that the ULDM proposal may not be able to alleviate the small-scale problems.
\end{abstract}


\section{Introduction}
Observational data of galactic rotation curves and hot gas in galaxy clusters reveal that dark matter exist. Besides, the data from the Cosmic Microwave Background \citep{Planck} and the large-scale structure observations \citep{Croft} have also provided a strong evidence for the dark matter existence. Most astrophysicists believe that dark matter consist of unknown particles which almost have no interaction with ordinary matter except gravity. Traditional proposals have suggested some hypothetical massive particles (dark matter mass $m>10$ MeV) to account for the dark matter. However, null detections of such particles have been obtained in both direct-detection experiments \citep{Aprile} and large hadron collider experiments \citep{Abecrcrombie}. Moreover, a large parameter space of these particles has been ruled out by gamma-ray and radio observations \citep{Ackermann,Chan,Chan2}. 

Recently, a class of dark matter called ultra-light dark matter (ULDM) has caught more attention. Some early studies have suggested the existence of light scalar or pseudo-scalar particles \citep{Peccei,Weinberg}. For example, a hypothetical particle called axion can help solve the CP-violation problem of the strong interaction \citep{Peccei} and it can be a possible kind of dark matter particle. Various studies show that ULDM particles may have self-interaction to form Bose-Einstein Condensate to mimic large dark matter structures in galaxies \citep{Zhang} or become dark matter superfluid \citep{Berezhiani}. Moreover, one important advantage is that the ULDM particles with mass $m \sim 10^{-22}$ eV can form small core structures in galaxies with radius $\sim$ kpc and they behave like cold dark matter (CDM) outside the cores \citep{Ferreira}. This particular property can simultaneously solve the small-scale problems, including the core-cusp problem \citep{deBlok}, the missing satellite problem \citep{Moore} and the too-big-to-fail problem \citep{Boylan}, encountered in the standard CDM model \citep{Ferreira}.

Many recent studies have constrained the mass range of ULDM particles using the data of dwarf galaxies. For example, by fitting the observed mass of Sculptor and Fornax dwarf galaxies, the upper limits of ULDM mass are $m_{22}<1.1$ \citep{Marsh} and $m_{22}<0.4$ \citep{Gonzalez} respectively, where $m_{22}=m/(10^{-22}~{\rm eV})$. Also, by fitting the core masses of the Draco and Sextans dwarf galaxies, one can get $m_{22}=0.8^{+0.5}_{-0.3}$ and $m_{22}=6^{+1}_{-2}$ \citep{Chen,Ferreira}. Another study considering the Draco II and Triangulum II dwarf galaxies conclude $m_{22} \sim 3.7-5.6$ \citep{Calabrese}. More recent studies using the data of the Milky Way \citep{Maleki,Li} and high-redshift galaxies \citep{Kulkarni} generally obtain $m_{22} \sim 1$. In view of these constraints, we can see that the possible ranges or limits of ULDM mass are not quite consistent with each other. 

In fact, many of these studies are model-dependent and rely on some unjustified assumptions. Generally speaking, constraining the ULDM mass using the central dark matter core properties requires the information of the total dark matter halo mass $M_h$. The uncertainties of $M_h$ are usually very large and the determination of $M_h$ is somewhat model-dependent. For instance, most studies assume the Navarro-Frenk-White (NFW) profile to describe the ULDM density profile outside the soliton core \citep{Marsh,Chen,Gonzalez}. However, observations of many galaxies seem to suggest an isothermal dark matter profile (or a cored isothermal dark matter profile) rather than the NFW dark matter profile \citep{Spano,Velander,Grillo}. Such a difference would significantly affect the estimated $M_h$. Also, the cutoff radius of a dark matter halo following the assumption of the NFW or isothermal profile is not very clear. The cutoff radius is usually assumed when the average dark matter density is equal to a certain factor (e.g. 200 times or 500 times) of the cosmological critical density, which may not be justified to precisely define the actual value of $M_h$. On the other hand, \citet{Safarzadeh} show that by relaxing the assumption of the slope constraint from classical dwarf galaxies, the allowed mass range would be significantly changed. The slope constraint used also depends on the dark matter profile assumed outside the core. Therefore, many previous conclusions are model-dependent and suffer from a large systematic uncertainty. We need to formulate a less model-dependent framework to give a more robust constraint of ULDM mass.

In this article, we formulate a model-independent method to constrain the ULDM mass range. In this theoretical framework, we only rely on the results of the ULDM numerical simulations without any presumptions of specific dark matter profile or halo mass function. By examining the data of the Spitzer Photometry \& Accurate Rotation Curves (SPARC), we discover that the rotation curve data of 5 galaxies can give meaningful and stringent constraints for ULDM. In particular, one of the galaxies ESO563-G021 can conservatively exclude a mass range $m_{22}=0.14-3.11$, which is a crucial range for the ULDM proposal to explain the small-scale problems in the standard CDM model. 

\section{The theoretical framework}
Assume that all dark matter are ULDM particles. Numerical simulations of the ULDM show that the density profile of the innermost central region of the dark matter halos at redshift $z=0$ follows \citep{Schive,Safarzadeh}
\begin{equation}
\rho_{\rm DM}=\frac{1.9(10m_{22})^{-2}r_c^{-4}}{[1+9.1\times 10^{-2}(r/r_c)^2]^8}10^9~M_{\odot}~{\rm kpc}^3,
\end{equation}
where $r_c$ is the core radius at which the density drops to one-half its peak value. Based on this density profile, we can get the enclosed core mass
\begin{equation}
M_c \approx {1}{4}\left(\frac{M_h}{M_{\odot}} \right)^{1/3}(4.4 \times 10^7m_{22}^{-3/2})^{2/3}M_{\odot},
\end{equation}
and the core radius
\begin{equation}
r_c \approx 1.6m_{22}^{-1} \left(\frac{M_h}{10^9M_{\odot}} \right)^{-1/3}~{\rm kpc}
\end{equation}
in terms of the ULDM mass $m_{22}$ and the total halo mass $M_h$ \citep{Schive,Safarzadeh}.

Since $M_h>10^{10}M_{\odot}$ for most of the galaxies, if we have $m_{22} \sim 1$, the core radius should be $r_c \sim 1$ kpc. Therefore, by examining the galactic observed data $\sim 1$ kpc from the galactic centers, we can get a constrained parameter space for $m_{22}$ and $M_h$. Consider a data point $(r_i,M_i)$ obtained from the rotation curve data of a galaxy, where $r_i$ and $M_i$ are the radius and the enclosed dark matter halo mass at $r_i$ from the galactic center. Since the dark matter halo mass profile $M_{\rm DM}(r)$ is an increasing function, for $r_i>r_c$, we must have $M_i \ge M_c$. Therefore, it is impossible to have $M_i<M_c$ for $r_i>r_c$. Thus, for a certain parameter space ($m_{22},M_h$) such that $M_i<M_c$ and $r_i>r_c$, that parameter space would be the excluded region (see Fig.~1).  

In order to get a constrained mass range of $m_{22}$ only, we need to constrain the value of $M_h$. Calculating the range of $M_h$ is usually model-dependent, like assuming the NFW profile or a certain halo mass function. Nevertheless, we can get a model-independent and conservative lower limit of $M_h$ by analyzing the outermost rotation curve data of a galaxy. For example, if the outermost rotation curve data point of a galaxy gives $M_j$ (the minimum total observed dark matter halo mass), the parameter space of $M_h<M_j$ is excluded. From the parameter space graph $(m_{22},M_h)$, if the position of the horizontal line $M_h=M_j$ is higher than the intersection point of the two solid lines (e.g. $(m_{22}',M_h')$ in Fig.~1), the excluded regions based on the data ($r_i,M_i$) and $M_j$ would separate the allowed parameter space of ($m_{22},M_h$) into two discrete regions (because the region below the horizontal line $M_j=M_h$ is excluded). The range of $m_{22}$ between the two separated regions would be the excluded $m_{22}$ range. 

To get the analytic excluded mass range of $m_{22}$ from the data $r_i$, $M_i$ and $M_j$, we can solve Eq.~(2) and Eq.~(3) directly by putting $M_i=M_c$ and $r_i=r_c$. Therefore, we get
\begin{equation}
M_i=3.1 \times 10^4 \left(\frac{M_h}{M_{\odot}} \right)^{1/3}m_{22}^{-1}M_{\odot},
\end{equation}
and
\begin{equation}
r_i=1600m_{22}^{-1}\left(\frac{M_h}{M_{\odot}}\right)^{-1/3}~{\rm kpc}.
\end{equation}
Since $M_h<M_j$ is excluded, we set $M_j=M_h$ to get the excluded $m_{22}$ range. Therefore, the analytic excluded $m_{22}$ range is
\begin{equation}
\frac{1600(M_j/M_{\odot})^{-1/3}}{(r_i/1~\rm kpc)}<m_{22}<\frac{3.1\times 10^4(M_j/M_{\odot})^{1/3}}{(M_i/M_{\odot})}.
\end{equation}

\section{Data analysis}
We consider the SPARC data in \citet{Lelli} to constrain $m_{22}$. The data consist of high-quality rotation curves from 175 nearby galaxies and include four velocity components at different radii $r$: total rotational velocity $V_{\rm tot}$, gas disk $V_{\rm gas}$, stellar disk $V_{\rm disk}$ and bulge $V_{\rm bul}$ \citep{Lelli}. The total enclosed mass $M(r)=rV_{\rm tot}^2/G$ is a sum of the enclosed dark matter halo mass $M_{\rm DM}(r)$ and the baryonic mass $M_{\rm bar}(r)$. Therefore, the dark matter halo mass can be obtained by \citep{Lelli}
\begin{equation}
M_{\rm DM}(r)=\frac{r}{G}\left[V_{\rm tot}^2-|V_{\rm gas}|V_{\rm gas}-\Upsilon_{\rm disk}|V_{\rm disk}|V_{\rm disk}-\Upsilon_{\rm bul}|V_{\rm bul}|V_{\rm bul} \right],
\end{equation}
where $\Upsilon_{\rm disk}$ and $\Upsilon_{\rm bul}$ are the stellar mass-to-light ratios for the disk and bulge components respectively. By examining the first few data points (i.e. at small radii $r$) and using Eq.~(7), we can obtain the data points $(r_i,M_i)$ for different galaxies. However, the actual values of $\Upsilon_{\rm disk}$ and $\Upsilon_{\rm bul}$ are not very well-known for different galaxies. To make a conservative calculations (i.e. maximizing the $M_{\rm DM}$), we take $\Upsilon_{\rm disk}=\Upsilon_{\rm bul}=0$ so that we get the possible maximum dark matter halo mass $M_{\rm DM,max}(r_i)=r_i(V_{\rm tot}^2-V_{\rm gas}|V_{\rm gas}|)/G \equiv M_i$ for different galaxies.

On the other hand, we need to calculate the minimum total dark matter halo mass $M_h$ for conservative calculations. We select the outermost rotation curve data point (at $r_j$) for each galaxy and use Eq.~(7) to calculate $M_j \equiv M_{\rm DM}(r_j)$. Since the bulge component is nearly negligible at the outermost region of a galaxy, we can safely assume $\Upsilon_{\rm bul}=0$. For the disk component, we take the conservative maximum value $\Upsilon_{\rm disk}=0.7$ \citep{Lelli} to obtain the minimum total dark matter halo mass $M_{\rm DM,min}(r_j)=M_j$ for our standard analysis. Generally speaking, increasing the value of the maximum $\Upsilon_{\rm disk}$ would give less stringent constraints for $m_{22}$. Beside the conservative value of $\Upsilon_{\rm disk}=0.7$, we will also obtain the limits of $m_{22}$ for different possible values of $\Upsilon_{\rm disk}$.

After checking with the data of the whole SPARC sample, we find that there are five galaxies (DDO161, ESO563-G021, IC4202, NGC1090, NGC6015) which have appropriate $(r_i,M_i)$ and $M_j$ that can separate the allowed parameter space of ($m_{22},M_h$) into two discrete regions. By using the analytic formula in Eq.~(6), we can get the excluded range of $m_{22}$ for each galaxy. The corresponding data and the excluded ranges of $m_{22}$ are shown in Table 1. In fact, the observed values of the total rotational velocity for each galaxy have the $1\sigma$ error bars $\Delta V_{\rm tot}$. The errors mainly come from the fitting procedures and the errors due to considering of the approaching and receding sides of the disk \citep{Lelli}. In our analysis, we have included the corresponding $1\sigma$ limit for each data point in determining the values of $M_i$ and $M_j$ (i.e. using the values of $V_{\rm tot}+\Delta V_{\rm tot}$ to determine $M_i$ and $V_{\rm tot}-\Delta V_{\rm tot}$ to determine $M_j$). Moreover, note that the gas velocities can be negative in the innermost regions because the gas distribution may have a significant central depression and the material in the outer regions could exert a stronger gravitational force than that in the inner parts \citep{Lelli}. In our five sample galaxies, both NGC1090 and NGC6015 have a negative velocity gas component in the innermost regions. We have considered this factor in our analysis, although the overall effect of the negative velocity gas component in these two galaxies is about 1\% only. Since the data points in \citet{Lelli} are continuous in $r$, we can combine the excluded ranges of $m_{22}$ obtained from the same galaxy and get the overall excluded ranges. Among the results in Table 1, the constraints obtained by the ESO563-G021 galaxy are the most stringent. The excluded mass range $m_{22}=0.14-3.11$ basically covers the excluded ranges based on the other four galaxies. 

In Fig.~2, we show the graphical analysis of the data from the ESO563-G021 galaxy. There are two data points ($r_i,M_i$) which can separate the allowed $m_{22}$ into two discrete regions (i.e. two pairs of lines are plotted). It has a relatively large minimum total dark matter halo mass $M_j$ so that the excluded range (shaded region) is somewhat wider. We also show the similar graphical analyses for the other four galaxies in Fig.~3. If we assume a larger value of $\Upsilon_{\rm disk}$, the value of $M_j$ for each galaxy would decrease and the corresponding limits would be less stringent. We show the corresponding excluded ranges of $m_{22}$ for three other values of $\Upsilon_{\rm disk}$ in Table 2. 

Note that some inner regions ($\le 0.5$ kpc) of our target galaxies are considered in our analysis. In such inner regions, the radial velocity component (or radial velocity dispersion) $v_r$ may dominate the velocity of the dark matter tracer's orbit so that the total enclosed mass calculated by the rotational speed may not be correct. Nevertheless, except the IC4202 galaxy, the other four galaxies do not have any spherical bulge component. Therefore, the effect of $v_r$ would be smaller in these four galaxies. In fact, the effect of the radial velocity component can be characterized by the anisotropy parameter $\beta=1-(v_t/v_r)^2$, where $v_t$ is the tangential velocity component. From the Jeans equation, the total enclosed mass is $M(r)=rv_r^2(\gamma_*+\gamma_r-2\beta)/G$, where $\gamma_*$ and $\gamma_r$ are the gradient of the stellar mass distribution and the gradient of $v_r$ respectively \citep{Wolf}. For the bulgeless galaxies at small $r \le 0.5$ kpc, the disk components with disk scale radius $r_d \sim 5-10$ kpc dominate the stellar distribution. For the region $r/r_d \le 0.1$, the disk stellar density is almost constant so that the stellar density gradient $\gamma_* \sim r/r_d$ and the value of $\gamma_r$ are relatively small. Therefore, the value of $M(r)$ is dominated by the term $2r(v_t^2-v_r^2)/G$. However, the contribution of the $v_r$ component is negative, which would give a smaller total enclosed mass. As a result, the enclosed dark matter halo mass calculated based on our data without considering the $v_r$ component would be overestimated so that the excluded $m_{22}$ ranges obtained are more conservative. Nevertheless, the effect of $v_r$ in IC4202 galaxy may be relatively larger because it has a bulge component which may have a larger value of $\gamma_*$ at small $r$. 

\begin{figure}
 \includegraphics[width=140mm]{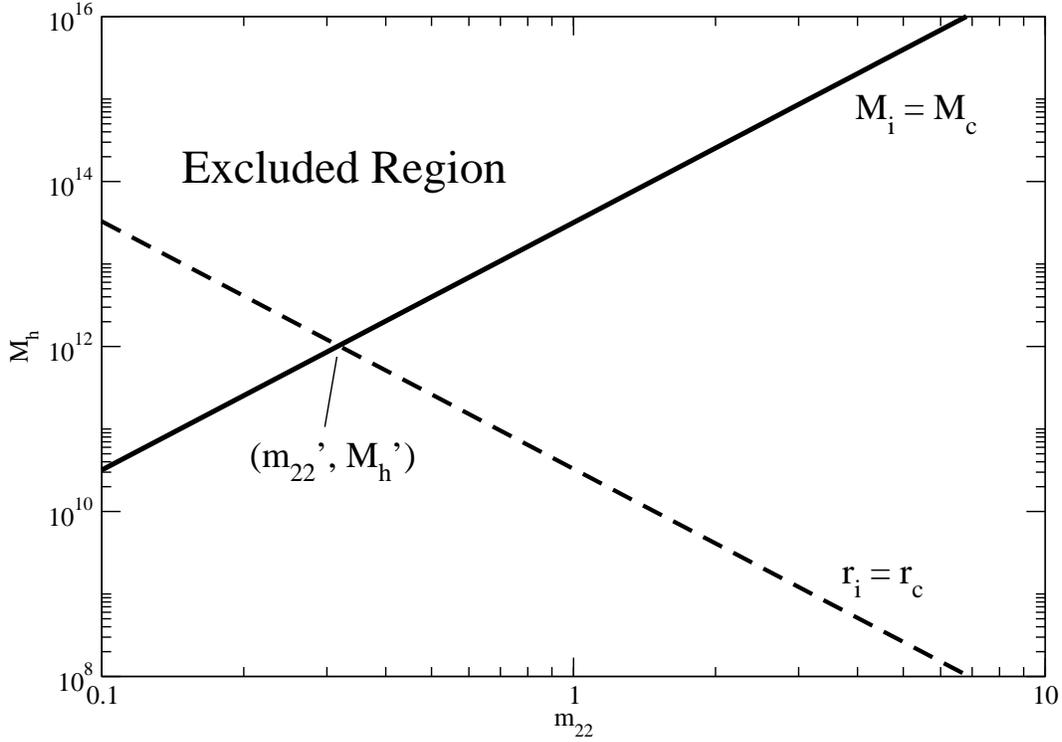}
\caption{The solid and dashed lines represent the relations $M_i=M_c$ and $r_i=r_c$ respectively. Here, we arbitrary set $M_i=10^9M_{\odot}$ and $r_c=0.5$ kpc for illustration. The upper region bounded by the solid and dashed lines is the excluded parameter space of $(m_{22},M_h)$. The $(m_{22}',M_h')$ is the intersection point of the solid and dashed lines. The unit for $M_h$ is in solar mass.}
\label{Fig1}
\vskip 10mm
\end{figure}

\begin{figure}
 \includegraphics[width=140mm]{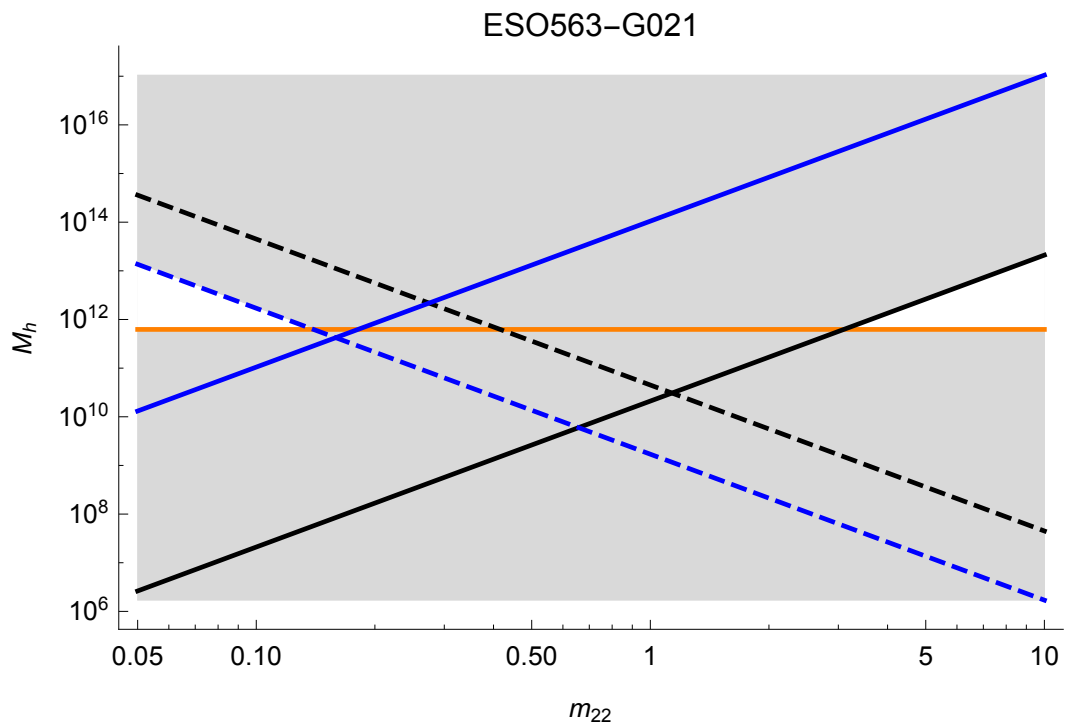}
\caption{The blue and black solid lines represent the relation $M_i=M_c$ (two data points) while the blue and black dashed lines represent the relation $r_i=r_c$ (two data points) for the ESO563-G021 galaxy. The orange solid horizontal line represents the lower limit of $M_h=M_j$ (assuming $\Upsilon_{\rm disk}=0.7$). The gray region is the excluded parameter space of $(m_{22},M_h)$. The unit for $M_h$ is in solar mass.}
\label{Fig2}
\vskip 10mm
\end{figure}

\begin{figure}
\vskip 10mm
 \includegraphics[width=70mm]{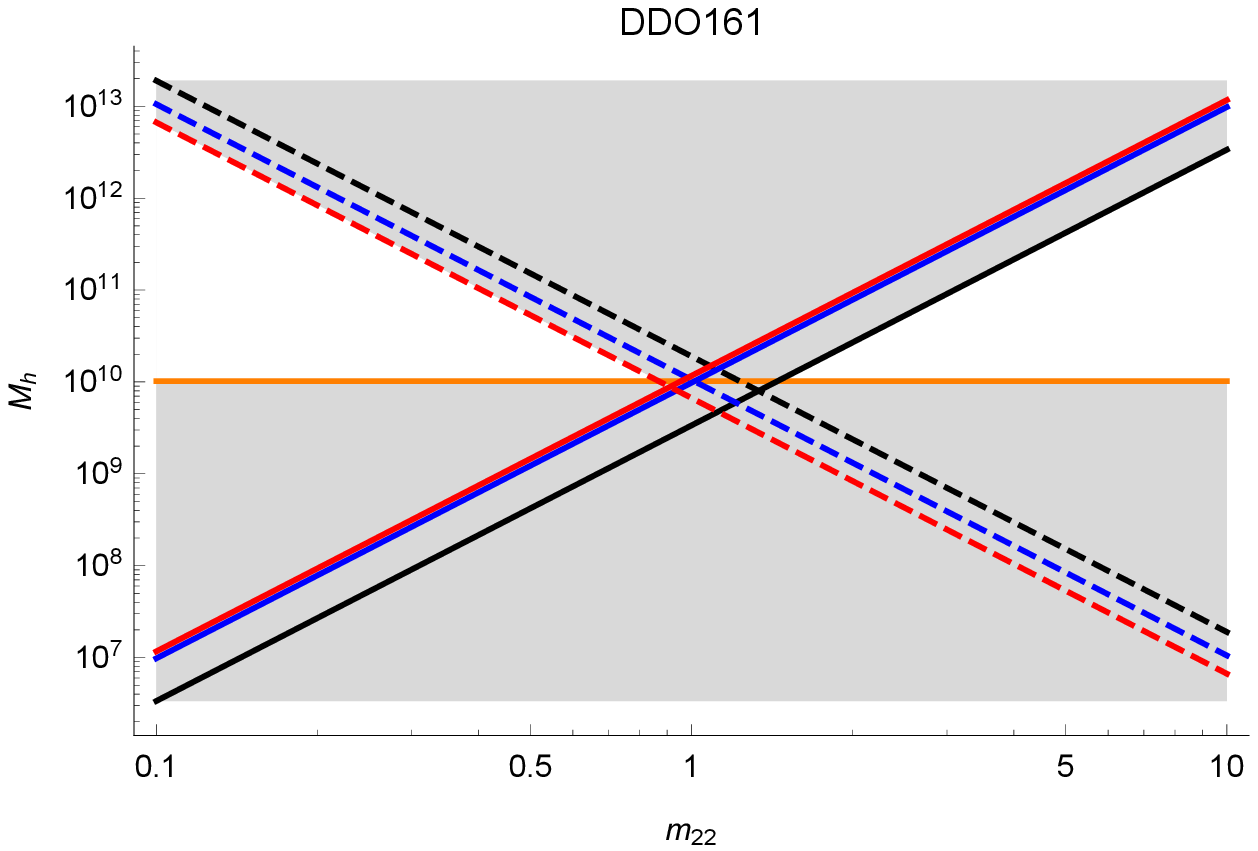}
~~ \includegraphics[width=70mm]{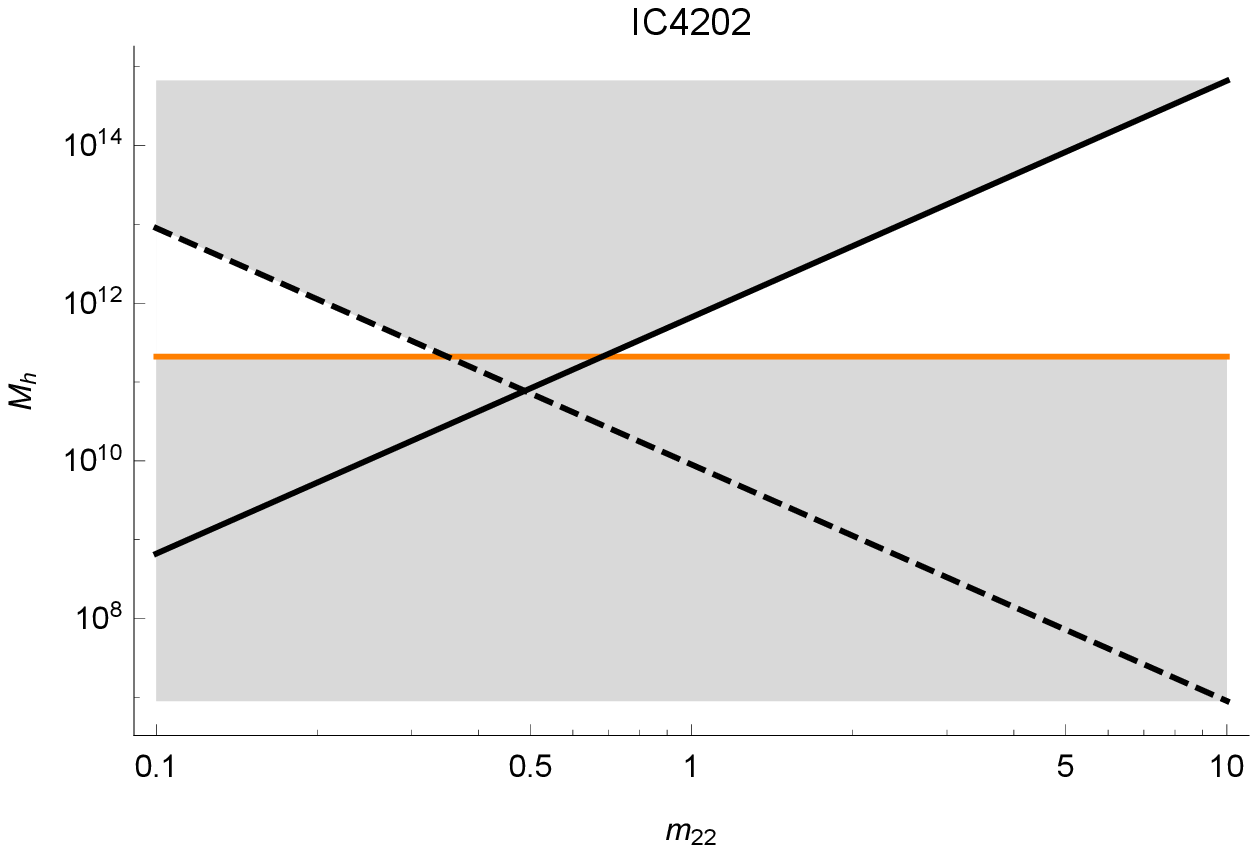} \newline
\newline
\newline
 \includegraphics[width=70mm]{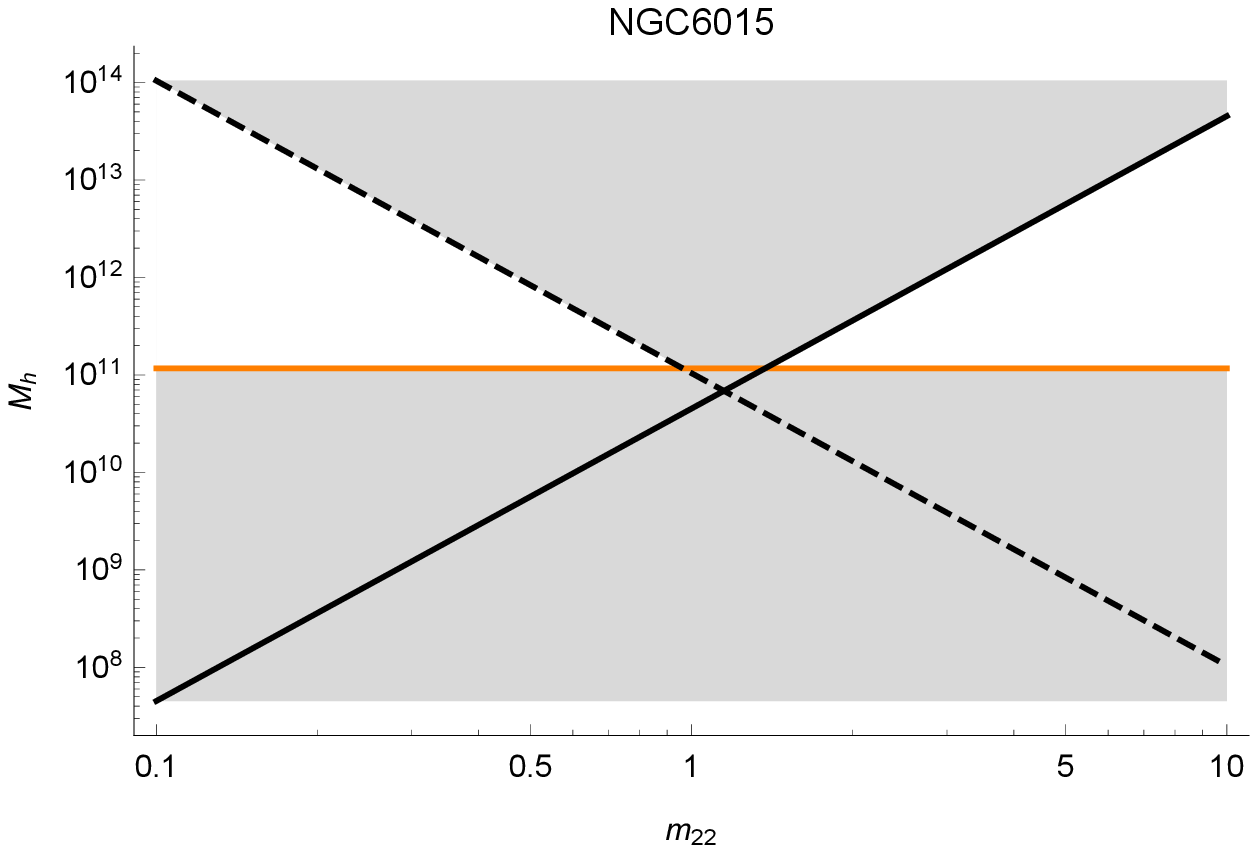}
~~ \includegraphics[width=70mm]{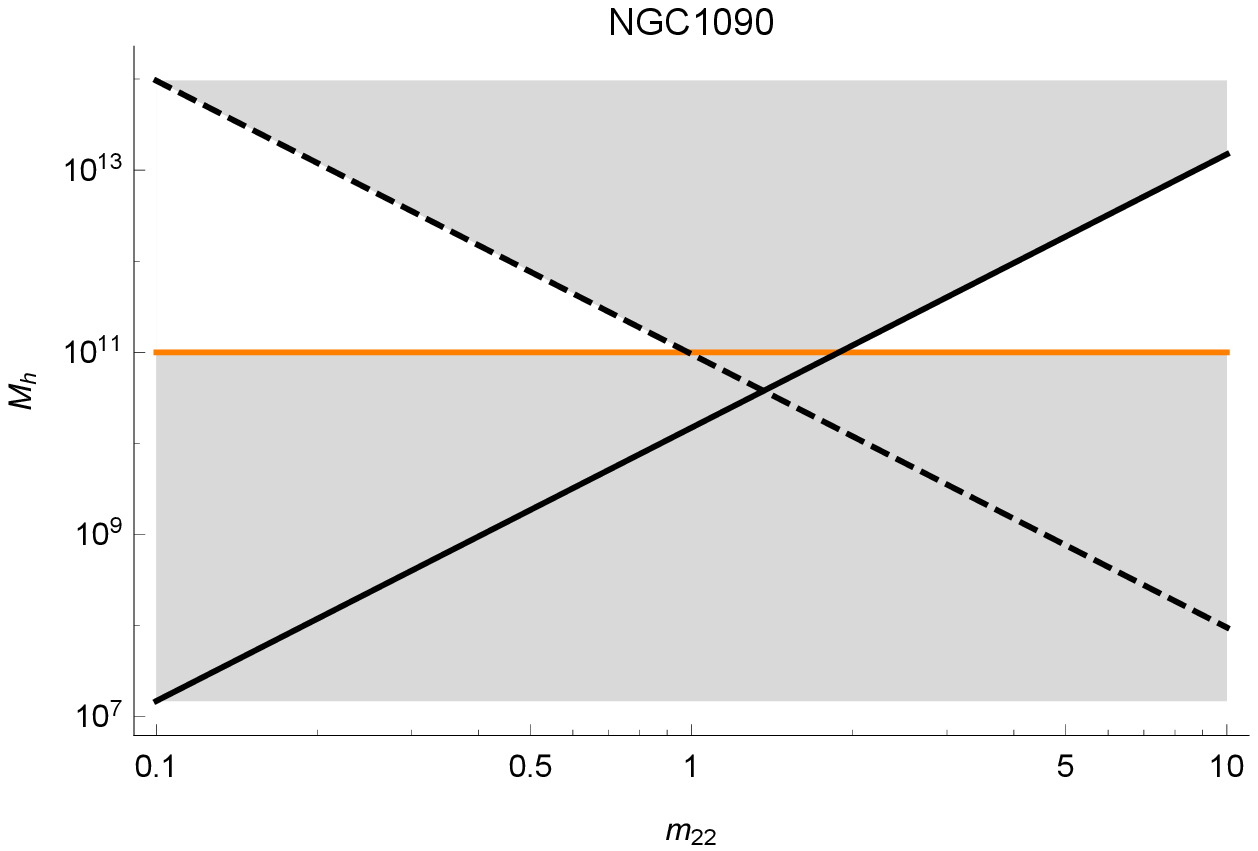}
 \caption{The blue, red and black solid lines represent the relation $M_i=M_c$ while the blue, red and black dashed lines represent the relation $r_i=r_c$ for four other galaxies DDO161, IC4202, NGC6015 and NGC1090 (there are three data points for DDO161 galaxy). The orange solid horizontal lines represent the lower limits of $M_h=M_j$ (assuming $\Upsilon_{\rm disk}=0.7$). The gray regions are the excluded parameter space of $(m_{22},M_h)$. The unit for $M_h$ is in solar mass.}
\label{Fig3}
\vskip 10mm
\end{figure}

\section{Discussion}
In this article, we have described a new and simple model-independent analysis to constrain $m_{22}$. We can write the excluded mass range of $m_{22}$ analytically in terms of the data of galactic rotation curves $r_i$, $M_i$ and $M_j$. Using the SPARC data, we find that there are five galaxies which can give stringent constraints for $m_{22}$.  Note that besides the five chosen galaxies, all the remaining galaxies in the SPARC sample cannot give any meaningful constraints based on our method because $M_j \le M_h'$ in these galaxies. In other words, the total enclosed mass content manifested by the outermost data in these galaxies is not large enough to give any useful limits (i.e. no excluded ranges of $m_{22}$ can be obtained). 

The overall excluded mass range is $m_{22}=0.14-3.11$ in our standard analysis, assuming a conservative upper limit of $\Upsilon_{\rm disk}=0.7$. This excluded mass range is conservative as we have maximized the possible dark matter halo mass at small $r$ by setting $\Upsilon_{\rm disk}=\Upsilon_{\rm bul}=0$ and minimized the possible total dark matter halo mass $M_h$ at large $r$ by putting a maximum value of $\Upsilon_{\rm disk}=0.7$. The actual excluded mass range should be larger as the actual value of $M_h$ is larger than the value of $M_j$ in these galaxies. Moreover, the dark matter halo mass is directly probed from the rotation curves, which does not have any presumption of specific dark matter density profile. The only assumption made in this analysis is the density core formed based on the numerical simulations (i.e. Eq.~(1)). 

In fact, the gravity and the baryonic feedback of the baryonic components may have significant impact on the resultant dark matter density profile. Recent numerical simulations involving baryons have examined how baryons affect the dark matter density core formed \citep{Mocz,Veltmaat}. In particular, \citet{Mocz} show that the dark matter structure at early times is largely unaffected by the effects of baryonic feedback and reionization. On the other hand, a more recent study shows that the dark matter core formed by the ULDM with baryons has a higher core density than that of the ULDM-only simulations at redshift $z=4.0$ \citep{Veltmaat}. If it is also true for $z=0$, this result would give a much tighter constraint for the upper limit of $m_{22}$ because it would give a larger $M_c$ in general. Therefore, our constraints based on the ULDM-only simulations would be more conservative. Besides the baryonic effects, the non-linear dynamical effects of the large-volume cosmological simulations on the ULDM density profile might also be important. However, due to the limited resolution, the effect is still uncertain and further investigations are required to understand the actual effect \citep{May}. Moreover, although numerical simulations show that the dark matter density behaves like the NFW profile in the outer region \citep{Schive}, we do not assume any specific dark matter density profile in our analysis because the baryonic effects have not been precisely considered in the numerical simulations. We have probed the dark matter halo profiles from the observed rotation curves, which are model-independent and more reliable. 

The most important feature of the ULDM model is that it can explain the core-like structures in galaxies and behave like CDM outside the central core region if $m_{22} \sim 1$ \citep{Ferreira}. Furthermore, it can also address the missing satellite problem and the too-big-to-fail problem \citep{Ferreira}. Therefore, the ULDM model with $m_{22} \sim 1$ can simultaneously solve the small-scale problems encountered by the CDM model and agree with the large-scale properties manifested by the CDM model (e.g. Ly$\alpha$ spectrum). However, the conservative excluded mass range $m_{22}=0.14-3.11$ obtained from our analysis has covered a crucial mass range that would significantly challenge the biggest advantage of the ULDM model. If $m_{22} \ge 3.11$, then the Jeans mass (minimum mass of substructure created) would be $\le 10^7M_{\odot}$, which means that halos with mass $\sim 10^7-10^8$ are still highly created. Therefore, the missing satellite problem and the too-big-to-fail problem would not be alleviated. Also, for a normal dark matter halo mass $M_h \ge 10^{10}M_{\odot}$, the core size would be less than 0.25 kpc if $m_{22} \ge 3.11$. Such a core size would be too small to account for the observed size, which may range from $\sim 0.3-10$ kpc \citep{Burkert}. Also, \citet{Burkert} has recently suggested that the fuzzy dark matter model (a kind of ULDM model) cannot account for the core formation in galaxies because the observed constant column density for galaxies is in disagreement with the ULDM simulation prediction. Therefore, our results seem to be consistent with this suggestion and would further wash out this most important advantage of the ULDM model. Nevertheless, the small-scale problems can still be explained by other possible mechanisms, such as dark matter self-interaction \citep{Spergel,Chan3} and baryonic feedbacks \citep{Pontzen}.

Generally speaking, our results are consistent with some recent bounds obtained. For example, \citet{Safarzadeh} suggest that $m_{22}$ should be larger than 6 if the slope constraints from classical dwarf galaxies are relaxed. Also, using the data of the Milky Way galaxy, \citet{Maleki} and \citet{Li} suggest $m_{22}=25^{+36}_{-20}$ and $m_{22} \sim 2-7$ respectively. Further investigation might put focus on $m_{22}>3$ or $0.01 \le m_{22} \le 0.1$ as Cosmic Microwave Background data generally allow $m_{22} \ge 0.01$ \citep{Hlozek}. However, these ranges of $m_{22}$ would not be able to alleviate the small-scale problems of the CDM model.

\begin{table}
\caption{The data derived from the observed rotation curves in \citet{Lelli} and the corresponding excluded mass ranges for the five galaxies (assuming $\Upsilon_{\rm disk}=0.7$).}
 \label{table1}
 \begin{tabular}{@{}lcccc}
  \hline
  Galaxy & $M_j$ ($M_{\odot}$) & $r_i$ (kpc) & $M_i$ ($M_{\odot}$) & excluded $m_{22}$ \\
  \hline
  \hline
  DDO161 & $1.02\times 10^{10}$ & 0.60 & $4.67\times 10^7$ & 1.23-1.46  \\
         &                      & 0.73 & $6.67\times 10^7$ & 1.01-1.02  \\
         &                      & 0.85 & $7.09\times 10^7$ & 0.87-0.96  \\
  \hline
  ESO563-G021 & $6.25\times 10^{11}$ & 0.45 & $8.60\times 10^7$ & 0.42-3.11  \\
              &                      & 1.34 & $1.47\times 10^9$ & 0.14-0.18  \\
  \hline
  IC4202 & $2.10 \times 10^{11}$ & 0.77 & $2.72 \times 10^8$ & 0.35-0.68  \\
  \hline
  NGC1090 & $9.97 \times 10^{10}$ & 0.35 & $7.67 \times 10^7$ & 0.99-1.89  \\
  \hline
  NGC6015 & $1.17\times 10^{11}$ & 0.34 & $1.11\times 10^8$ & 0.96-1.36  \\
  \hline
 \end{tabular}
\end{table}

\begin{table}
\caption{The excluded ranges of $m_{22}$ for the five galaxies with different $\Upsilon_{\rm disk}$.}
 \label{table2}
 \begin{tabular}{@{}lcccc}
  \hline
  Galaxy & $\Upsilon_{\rm disk}=0.7$ & $\Upsilon_{\rm disk}=0.9$ & $\Upsilon_{\rm disk}=1.1$ & $\Upsilon_{\rm disk}=1.3$ \\
  \hline
  \hline
  DDO161 & 0.87-1.46 & 0.87-1.45 & 0.87-1.45 & 0.88-1.44  \\
  \hline
  ESO563-G021 & 0.14-3.11 & 0.15-2.99 & 0.15-2.86 & 0.16-2.72  \\
  \hline
  IC4202 & 0.35-0.68 & 0.37-0.65 & 0.40-0.60 & 0.44-0.55  \\
  \hline
  NGC1090 & 0.99-1.89 & 1.05-1.79 & 1.12-1.67 & 1.22-1.53  \\
  \hline
  NGC6015 & 0.96-1.36 & 0.98-1.34 & 1.00-1.31 & 1.02-1.28  \\
  \hline
 \end{tabular}
\end{table}

\begin{acknowledgements}
The work described in this paper was partially supported by a grant from the Research Grants Council of the Hong Kong Special Administrative Region, China (Project No. EdUHK 28300518).
\end{acknowledgements}


\begin{thebibliography}{}
\bibitem[Abecrcrombie et al. (2020)]{Abecrcrombie} Abecrcrombie D. {\it et al.}, 2020, Phys. Dark Univ. 27, 100371.
\bibitem[Ackermann et al. (2015)]{Ackermann} Ackermann M. {\it et al.}, 2015, Phys. Rev. Lett. 115, 231301.
\bibitem[Aprile et al. (2018)]{Aprile} Aprile E. {\it et al.}, 2018, Phys. Rev. Lett. 121, 111302.
\bibitem[Berezhiani \& Khoury (2016)]{Berezhiani} Berezhiani L. \& Khoury J., 2016, Phys. Lett. B 753, 639.
\bibitem[Boylan-Kolchin, Bullock \& Kaplinghat (2011)]{Boylan} Boylan-Kolchin M., Bullock J. S. \& Kaplinghat M., 2011, Mon. Not. R. Astron. Soc. 415, L40.
\bibitem[Burkert (2020)]{Burkert} Burkert A., 2020, Astrophys. J. 904, 161.
\bibitem[Calabrese \& Spergel (2016)]{Calabrese} Calabrese E. \& Spergel D. N., 2016, Mon. Not. R. Astron. Soc. 460, 4397.
\bibitem[Chan (2013)]{Chan3} Chan M. H., 2013, Mon. Not. R. Astron. Soc. 433, 2310.
\bibitem[Chan \& Leung (2017)]{Chan} Chan M. H. \& Leung C. H., 2017, Sci. Rept. 7, 14895.
\bibitem[Chan et al. (2019)]{Chan2} Chan M. H., Cui L., Liu J. \& Leung C. S., 2019, Astrophys. J. 872, 177.
\bibitem[Chen, Schive \& Chiueh (2017)]{Chen} Chen S.-R., Schive H.-Y. \& Chiueh T., 2017, Mon. Not. R. Astron. Soc. 468, 1338.
\bibitem[Croft et al. (1999)]{Croft} Croft R. A. C., Weinberg D. H., Pettini M., Hernquist L. \& Katz N., 1999, Astrophys. J. 520, 1.
\bibitem[de Blok (2010)]{deBlok} de Blok W. J. G., 2010, Adv. Astron. 2010, 789293.
\bibitem[Ferreira (2020)]{Ferreira} Ferreira E. G. M., arXiv:2005.03254.
\bibitem[Gonz\`alez-Morales et al. (2017)]{Gonzalez} Gonz\`alez-Morales A. X., Marsh D. J. E., Pe\~narrubia J. \& Ure\~na-L\`opez L. A., 2017, Mon. Not. R. Astron. Soc. 472, 1346.
\bibitem[Grillo (2012)]{Grillo} Grillo C., 2012, Astrophys. J. 747, L15.
\bibitem[Hlozek et al. (2015)]{Hlozek} Hlozek R., Grin D., Marsh D. J. E. \& Ferreira P. G., 2015, Phys. Rev. D 91, 103512.
\bibitem[Kulkarni \& Ostriker (2020)]{Kulkarni} Kulkarni M. \& Ostriker J. P., arXiv:2011.02116.
\bibitem[Lelli, McGaugh \& Schombert (2016)]{Lelli} Lelli F., McGaugh S. S. \& Schombert J. M., 2016, Astron. J. 152, 157.
\bibitem[Li, Shen \& Shive (2020)]{Li} Li Z., Shen J. \& Schive H.-Y., 2020, Astrophys. J. 889, 88.
\bibitem[Maleki, Baghram \& Rahvar (2020)]{Maleki} Maleki A., Baghram S. \& Rahvar S., 2020, Phys. Rev. D 101, 103504.
\bibitem[Marsh \& Pop (2015)]{Marsh} Marsh D. J. E. \& Pop A.-R., 2015, Mon. Not. R. Astron. Soc. 451, 2479.
\bibitem[May \& Springel (2021)]{May} May S. \& Springel V., 2021, arXiv:2101.01828.
\bibitem[Mocz et al. (2019)]{Mocz} Mocz P. {\it et al.}, 2019, Phys. Rev. Lett. 123, 141301.
\bibitem[Moore et al. (1999)]{Moore} Moore B. {\it et al.}, 1999, Astrophys. J. 524, L19.
\bibitem[Peccei \& Quinn (1977)]{Peccei} Peccei R. D. \& Quinn H. R., 1977, Phys. Rev. Lett. 38, 1440.
\bibitem[Planck Collaboration (2020)]{Planck} Planck Collaboration Aghanim N. {\it et al.}, 2020, Astron. Astrophys. 641, A6.
\bibitem[Pontzen \& Governato (2014)]{Pontzen} Pontzen A. \& Governato F., 2014, Nature 506, 171.
\bibitem[Safarzadeh \& Spergel (2020)]{Safarzadeh} Safarzadeh M. \& Spergel D. N., 2020, Astrophys. J. 893, 21.
\bibitem[Schive et al. (2014)]{Schive} Schive H.-Y., Liao M.-H., Woo T.-P., Wong S.-K., Chiueh T., Broadhurst T. \& Hwang W.-Y. P., 2014, Phys. Rev. Lett. 113, 1290.
\bibitem[Spano et al. (2008)]{Spano} Spano M., Marcelin M., Amram P., Carignan C., Epinat B. \& Hernandez O., 2008, Mon. Not. R. Astron. Soc. 383, 297.
\bibitem[Spergel \& Steinhardt (2000)]{Spergel} Spergel D. N. \& Steinhardt P. J., 2000, Phys. Rev. Lett. 84, 3760.
\bibitem[Velander, Kuijken \& Schrabback (2011)]{Velander} Velander M., Kuijken K. \& Schrabback T., 2011, Mon. Not. R. Astron. Soc. 412, 2665.
\bibitem[Veltmaat, Schwabe \& Niemeyer (2020)]{Veltmaat} Veltmaat J., Schwabe B. \& Niemeyer J. C., 2020, Phys. Rev. D 101, 083518.
\bibitem[Weinberg (1978)]{Weinberg} Weinberg S., 1978, Phys. Rev. Lett. 40, 223.
\bibitem[Wolf et al. (2010)]{Wolf} Wolf J., Martinez G. D., Bullock J. S., Kaplinghat M., Geha M., Munoz R. R., Simon J. D. \& Avedo F. F., 2010, Mon. Not. R. Astron. Soc. 406, 1220.
\bibitem[Zhang et al. (2018)]{Zhang} Zhang X., Chan M. H., Harko T., Liang S.-D. \& Leung C. S., 2018, Eur. Phys. J. C, 78, 346.
\end{thebibliography}
\end{document}